\documentclass[12pt,preprint]{aastex}

\usepackage{graphicx}
\usepackage{color}

\slugcomment{ApJLett}

\shorttitle{P/2013 R3}
\shortauthors{Jewitt et al.}

\begin{document}

\title{Disintegrating Asteroid P/2013 R3}
\author{David Jewitt$^{1,2}$, Jessica Agarwal$^3$, Jing Li$^{1}$, Harold Weaver$^4$,  Max Mutchler$^5$ and Stephen Larson$^6$
}
\affil{$^1$Department of Earth, Planetary and Space Sciences,
UCLA, 
595 Charles Young Drive East, 
Los Angeles, CA 90095-1567\\
$^2$Dept.~of Physics and Astronomy,
University of California at Los Angeles, \\
430 Portola Plaza, Box 951547,
Los Angeles, CA 90095-1547\\
$^3$ Max Planck Institute for Solar System Research, Max-Planck-Str. 2, 37191 Katlenburg-Lindau, Germany\\
$^4$ The Johns Hopkins University Applied Physics Laboratory, 11100 Johns Hopkins Road, Laurel, Maryland 20723  \\
$^5$ Space Telescope Science Institute, 3700 San Martin Drive, Baltimore, MD 21218 \\
$^6$ Lunar and Planetary Laboratory, University of Arizona, 1629 E. University Blvd.
Tucson AZ 85721-0092 \\
}

\email{jewitt@ucla.edu}

\begin{abstract}
Splitting of the nuclei of comets into multiple components has been frequently observed but, to date, no main-belt asteroid has been observed to break-up.  Using the Hubble Space Telescope, we find that main-belt asteroid P/2013 R3 consists of 10 or more distinct components, the largest  up to 200 m in radius (assumed geometric albedo of 0.05) each of which produces a coma and comet-like dust tail. A diffuse debris cloud with total mass $\sim$2$\times$10$^8$ kg further envelopes the entire system.  The velocity dispersion among the components, $\Delta V \sim$ 0.2 to 0.5 m s$^{-1}$, is comparable to the gravitational escape speeds of the largest members, while their extrapolated plane-of-sky motions suggest break-up between February and September 2013.   The broadband optical colors are those of a C-type asteroid. We find no spectral evidence for gaseous emission, placing model-dependent upper limits to the water production rate $\le$1 kg s$^{-1}$.   Breakup may be due to a rotationally induced structural failure of the precursor body.
\end{abstract}

\keywords{minor planets, asteroids: general --- minor planets, asteroids: individual (P/2013 R3) --- comets: general}

\section{INTRODUCTION}
Main-belt object P/2013 R3 (Catalina-Pan STARRS, hereafter ``R3'') was 
discovered on UT 2013 September 15 and 
announced on September 27 (Hill et al.~2013).  Its orbital semimajor axis, eccentricity and inclination are 3.033 AU, 0.273 and 0.90\degr, respectively, firmly establishing R3 as a member of the main asteroid belt, although its dusty appearance resembles that of a comet. The Tisserand parameter relative to Jupiter, $T_J$ = 3.18, is significantly larger than the nominal ($T_J$ = 3) dividing line separating dynamical comets ($T_J <$ 3) from asteroids ($T_J >$ 3, c.f.~Kresak 1980).  The combination of asteroid-like orbit and comet-like appearance together qualify R3 as an active asteroid (Jewitt 2012) or, equivalently, a main-belt comet (Hsieh and Jewitt 2006).    The  mechanism responsible for mass loss in the majority of such objects is unknown.  

In this brief report, we describe initial observations taken to establish the basic properties of this remarkable object.   
At the time of observation, R3 had just passed perihelion ($R$ = 2.20 AU) on UT 2013 August 05.

\section{OBSERVATIONS} 
We used the Keck 10-m  telescope on Mauna Kea, Hawaii with LRIS, the Low Resolution Imaging Spectrometer at an image scale of  0.135\arcsec~pixel$^{-1}$ (Oke et al.~1995).    Observations through Johnson-Cousins BVRI filters were internally calibrated using flat field  images of an illuminated spot on the inside of the observatory dome.  Seeing-limited image quality was variable in the range 0.6\arcsec~to 0.8\arcsec~full width at half maximum (FWHM).  Keck data (Figure \ref{images}) 
on 2013 October 01 and 02 revealed three distinct, co-moving components embedded in a dust envelope extending $>$30\arcsec~in the projected anti-solar direction (independently reported in a press release by J. Licandro et al.~from observations taken October 11 and 12).

On HST we used the WFC3 camera \citep{2012wfci.book.....D} whose 0.04\arcsec~pixels each correspond to about 41 km at the distance of R3. The Nyquist-sampled spatial resolution is $\sim$82 km.  All observations were taken using the very broad F350LP filter (4758\AA~FWHM) which has an effective wavelength of 6230\AA~on a solar-type (G2V) source.  From each orbit we obtained five exposures of 348 s duration and one of 233 s. The observational geometry is summarized in Table \ref{geometry}.  The images are shown in Figure (\ref{images}).

The integrated brightness of R3 was monitored using the Berkeley KAIT (Katzmann Automatic Imaging Telescope), located on Mt. Hamilton, California (Richmond et al.~1993).  This  automated, 0.6 meter diameter telescope is equipped with a 512$\times$512 pixel CCD camera (scale 0.8\arcsec~pixel$^{-1}$).  KAIT provided nightly sequences of 30 integrations each of 30 s duration through an R filter,  with image quality $\sim$2\arcsec~FWHM.  The images were shifted and combined to eliminate cosmic rays and to provide an improved signal-to-noise ratio.

\subsection{MORPHOLOGY AND DYNAMICS}
\begin{figure}
\epsscale{0.65}
\begin{center}
\plotone{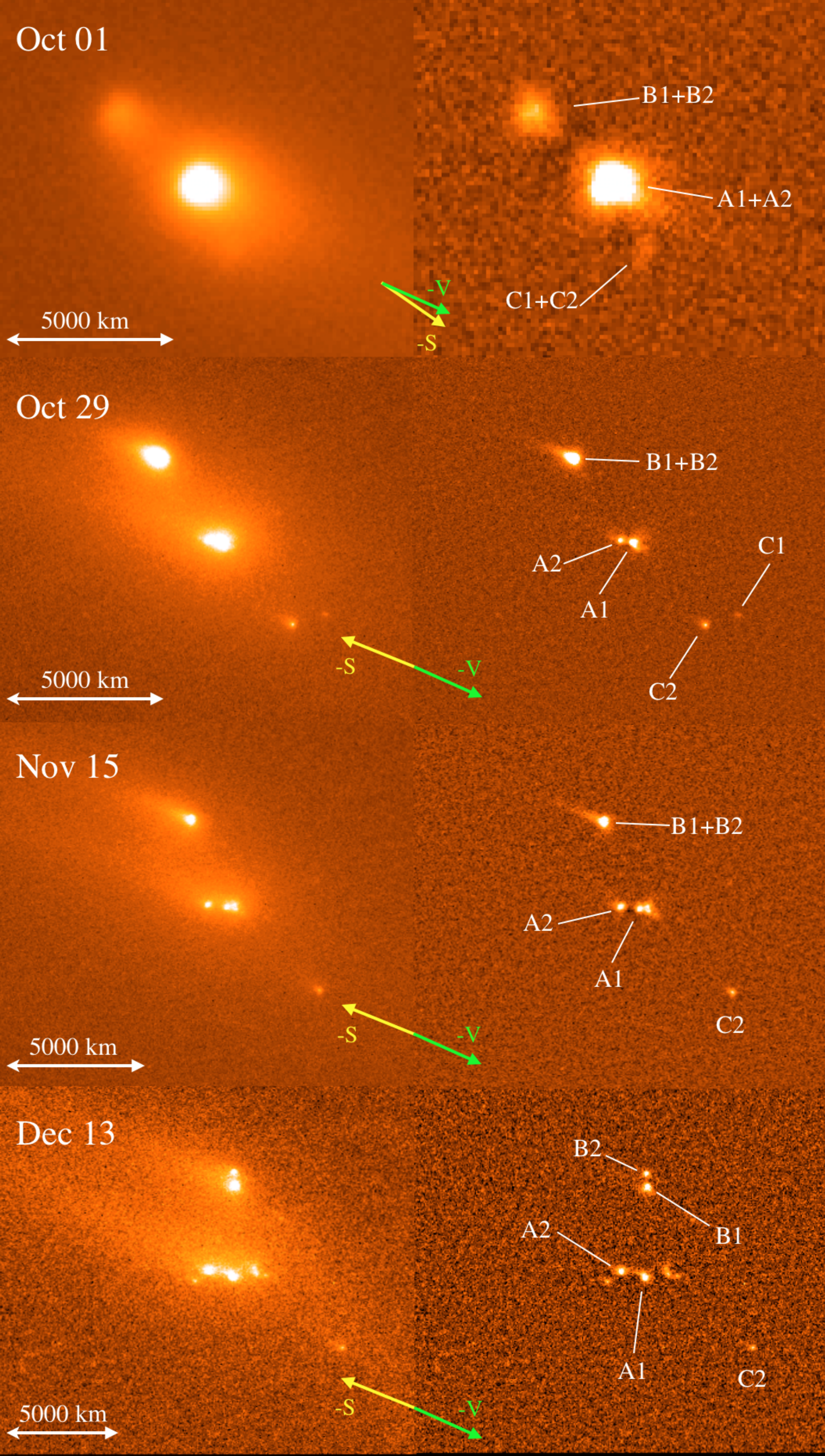}
\caption{Four epochs of R3 imaging from 2013 (c.f.~Table \ref{geometry}) shown as raw images (left column) and spatially filtered to suppress diffuse coma (right column).  October 01 data are from Keck, all the rest from HST. Each panel has North to the top, East to the left and has dimensions 14\arcsec~$\times$12\arcsec.  The projected anti-solar direction is shown by a yellow arrow marked ``-S''.  Projected negative velocity vector is indicated by a green arrow marked ``-V''.   Components discussed in the text are identified.  \label{images}
} 
\end{center} 
\end{figure}

Figure (\ref{images}) shows three groups of objects (labeled A, B and C) initially distributed along a line at position angle 40\degr,  corresponding neither to the projected orbit nor to the anti-solar direction. This is unlike split comets, where the major components are spread along the projected orbit (Ishiguro et al.~2009, Reach et al.~2009) and the minor ones along the anti-solar direction (Weaver et al.~2001, 2008).  The number of components (by December 13 we detect ten distinct components, most formed by the disintegration of A and B), the sky-plane separations between them, $L$ (scaled to 1 AU),  and their brightnesses all change with time. We measured, $L$ and $dL/dt$ for the components taken pair-wise, and calculated nominal ages, $\tau = L/(dL/dt)$.  The separations projected to zero over dates in the range 140 $\le DOY \le$ 270 (May to September 2013), where DOY is the day of  year  in 2013.  One object (component ``D'') not shown in Figure (\ref{images}) was detected  36.4\arcsec~from component B in position angle 241.9\degr,  on UT 2013 October 01.  The object was also identified at 37.3\arcsec~and 239.0\degr~in data from the Magellan telescope on UT October 29, kindly provided by Scott Sheppard and was reported in a press release by Licandro et al.  The age for object D is $\tau \sim$ 8 months (DOY 50).  We consider these estimates preliminary pending acquisition of further astrometric data from which proper orbits and possible non-gravitational accelerations can be constrained.

The position angles of the dust tails  (243.7\degr$\pm$0.5\degr~on October 01 and 243.2\degr$\pm$0.6\degr~October 28) correspond to discordant synchrone dates (September 04$\pm$7 and July 10$\pm$13, respectively). 

Each discrete component appears embedded in a dust coma having steep surface brightness gradients in the central arcsecond.  
We experimented with schemes to inwardly extrapolate the coma surface brightness, in order to isolate the  brightness contribution from the embedded nuclei.  However, we found these schemes to be critically dependent on the extrapolation method, yielding highly uncertain results. 
Here, we elect to present robust upper limits to the embedded nuclei based on photometry obtained within  circular projected apertures 0.2\arcsec~in radius with background subtraction from a contiguous annulus extending to 0.4\arcsec~(Table \ref{photometry}).  Large aperture  measurements were taken to assess the integrated dust cross-section.

Absolute magnitudes of the nuclei were computed from 

\begin{equation}
H_V = V - 2.5 \log_{10}\left[R^2 \Delta^2\right] + 2.5 \log_{10}[\Phi(\alpha)].
\label{inversesq}
\end{equation}

\noindent  Here, $\Phi(\alpha)$, is the ratio of the brightness at phase angle $\alpha$ to that at phase angle 0\degr, estimated from Bowell et al.~(1989).  We used the phase function parameter $g$ = 0.15 as applicable to C-type asteroids.   

The absolute magnitudes are related to the geometric albedo, $p_V$, and the radius, $r_e$ [km], of a circle having a scattering cross-section equal to that of the object by

\begin{equation}
r_e = \frac{690}{p_V^{1/2}} 10^{-H_V/5}.
\label{rn}
\end{equation}

\noindent We take $p_V$ = 0.05 to compute effective radii as listed in Table \ref{photometry}.  The largest components, A1, A2, B1 and B2, all have $r_e \sim$ 0.2 km. Because of dust contamination we only know that the nucleus radii are $r_n \le r_e$ and we cannot determine which, if any, of the components in R3 is the primary (mass-dominant) one.    However,  the photometric limits to $r_n$ are sufficient to show that mutual gravitational interactions are negligible.  The angular scale of the Hill sphere of a body having radius $r_n$ is $\theta_H \sim r_n/(3 R_{\odot})$, where $R_{\odot}$ is the radius of the Sun.  With $r_n$ = 200 m, we find $\theta_H \sim$ 0.02\arcsec~($\sim$20 km), which is beneath the resolution of HST. The components in Figure (\ref{images})  can be safely assumed to move independently.

\subsection{COLOR AND SPECTRUM}
We used LRIS images and a photometry aperture 6.0\arcsec~in radius to determine the global colors of the object on UT 2013 Oct 01. Photometric calibration was obtained from separate observations of nearby standard stars (Landolt 1992).  We obtained B-V = 0.66$\pm$0.04, V-R = 0.38$\pm$0.03 and R-I = 0.36$\pm$0.03 mag., while the solar colors in the same filters are B-V = 0.64$\pm$0.02, V-R = 0.35$\pm$0.01, R-I = 0.33$\pm$0.01 mag (Holmberg et al.~2006).  Evidently, R3 is a neutral C-type object, as are $\sim$50\% of asteroids beyond 3 AU  (DeMeo and Carry 2013).  Note that the scattering cross-section in our data is dominated by dust, not by the embedded parent nuclei.  

\begin{figure}
\epsscale{0.85}
\begin{center}
\includegraphics[width=0.95\textwidth, angle =0 ]{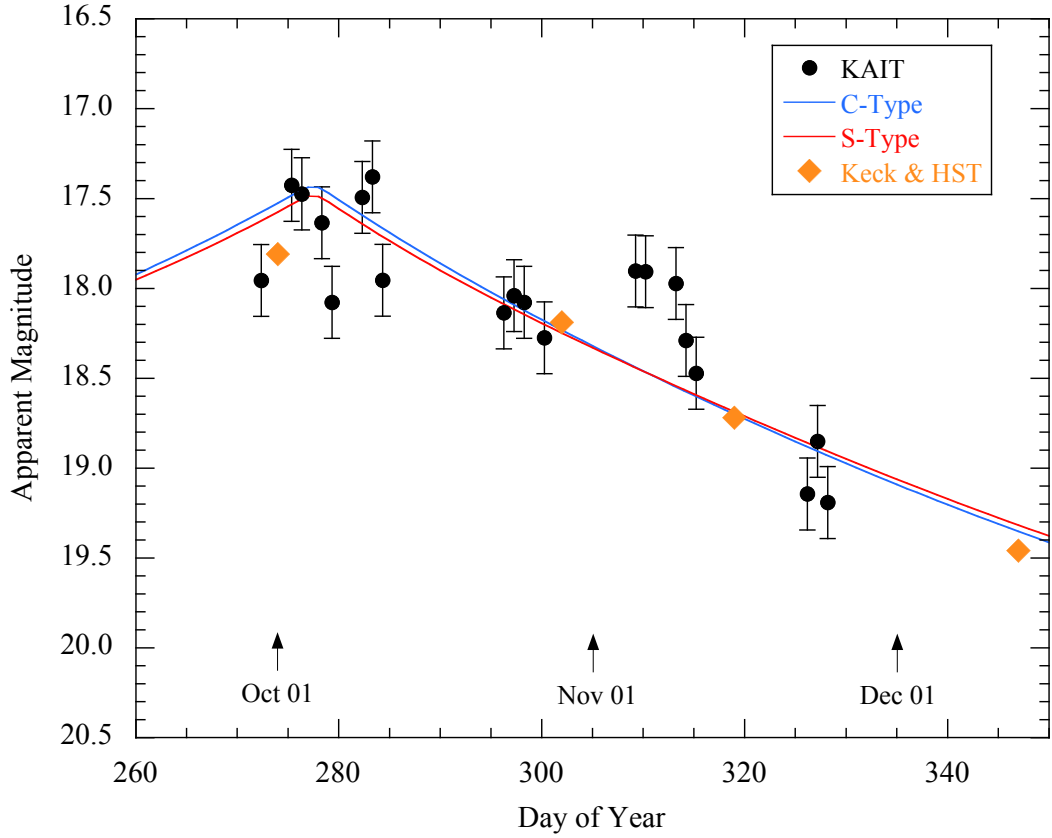}
\caption{Normalized reflection spectrum in the vicinity of the CN band (outlined in red dashed box).  Blue lines marked CW1 and CW2 denote regions of the spectrum used for continuum assessment. The 5$\sigma$ limit to CN emission is marked with a horizontal long-dashed line.  Residual H and K lines of calcium are marked.   \label{spectrum}
} 
\end{center} 
\end{figure}

We also used LRIS to obtain a spectrum of R3 on UT 2013 Oct 02, in search of gaseous emission bands.  The cyanide radical (CN) band-head at 3888\AA~(e.g.~A'Hearn et al.~1995) is the brightest optical line in comets and is used here as a proxy for water.  We obtained a 2000 s spectrum using a 1.0\arcsec~wide slit, oriented along position angle 50\degr~(the approximate tail axis) and a 400/3400 grism, obtaining a dispersion of 1.07\AA~pixel$^{-1}$ and a wavelength resolution of 7\AA, FWHM.  Wavelength calibration and spectral flat fields were obtained immediately following the R3 integrations. We also observed the G-type stars SA 115-271 and SA 93-101 for reference.  The former star provided better cancellation of the H and K lines of calcium at 3933\AA~and 3966\AA, and so we used this star to compute the reflectivity (Figure \ref{spectrum}). 

We extracted the spectrum from a 3.9\arcsec~long section of the  slit.  We determined a 3$\sigma$ upper limit to the flux density from CN by fitting the flanking continuum (wavelengths 3780\AA~to 3830\AA~on the blue side and 3900\AA~to 3950\AA~ on the red side) and taking into account the scatter in the data.  The resulting upper limit to the flux density is $f_{CN}$ = 1.3$\times$10$^{-16}$ erg cm$^{-2}$ s$^{-1}$.  


To calculate the production rate, $Q_{CN}$, we adopted a Haser model with parent and daughter scale lengths $\ell_p$ = 1.3$\times$10$^4$ km and $\ell_s$ = 2.1$\times$10$^5$ km, respectively, both at $R$ = 1 AU (A'Hearn et al.~1995) and a fluorescence efficiency at 1 AU, $g(1)$ = 10$^{-12.5}$ (Schleicher 2010).  We integrated the Haser model over the 1.0\arcsec$\times$3.9\arcsec~slit, projected to the distance of the object, and assumed a gas outflow speed of 500 m s$^{-1}$.  The non-detection of CN then corresponds to an upper limit $Q_{CN} <$ 1.2$\times$10$^{23}$ s$^{-1}$. The ratio of water to CN production rates in comets is $Q_{H_2O}/Q_{CN} \sim$360 (A'Hearn et al.~1995).  If this ratio applies to R3, then we deduce $Q_{H_2O} <$ 4.3$\times$10$^{25}$ s$^{-1}$, corresponding to $dM/dt <$ 1.2 kg s$^{-1}$.  At $R$ = 2.25 AU, a perfectly absorbing, spherical, dirty ice nucleus sublimating from the day-side would lose $F_s \sim$  2$\times$10$^{-5}$ kg m$^{-2}$ s$^{-1}$, in equilibrium.  The limit to the area of ice exposed on the nucleus is $A$ = $(dM/dt)/F_s  \sim$ 6$\times$10$^4$ m$^2$, corresponding to a hemisphere of radius $\sim$100 m.  However, the cometary ratio of $Q_{H_2O}/Q_{CN}$ might not apply to ice in a main-belt object and so the significance of the inferred $Q_{H_2O}$ is unclear.

\subsection{DUST PRODUCTION}
All images of R3 taken after October 01 show both a tail of months-old particles in the direction of the negative velocity vector (west of the nuclei), and tails of new material in the anti-solar direction (east). The simultaneous presence of tails in both directions is a strong indication that activity is on-going ($>$2-3 months). 

In order to study the distribution of the dust, we calculated synchrones (locations of particles ejected at constant time) and syndynes (locations of particles having fixed dimensionless radiation pressure factor, $\beta$) for each observation date (Finson and Probstein, 1968).   The position angles of the eastern tails of different fragments  correspond to synchrone dates from late September to early October.  The largest detectable value of $\beta$ was 0.7, 0.02, and 0.007 at the three HST epochs, corresponding to radii of 1, 30, and 100 $\mu$m. 
The smallest of these particles  were quickly removed by radiation pressure and therefore are not observed in the later images. 

Each component of R3 resembles a typical active comet, with a roughly spherical coma being blown back by radiation pressure. This suggests that the initial velocity of the particles was not negligible and, therefore, that the Finson-Probstein approach is not strictly valid. 
The sunward extension of the largest coma (A2) was about $s$ = 2000\,km. 
Following Jewitt et al.~(2011), we connect $s$, the turn-around distance in the sunward direction, to $u$, the initial sunward particle speed from

\begin{equation}
u^2 = 2 \beta g_{\odot} s,
\end{equation}

\noindent where  $g_{\odot}$ is the gravitational acceleration towards the Sun. Substituting for $g_{\odot}$ we obtain

\begin{equation}
\beta = \frac{u^2 R^2}{2 G M_{\odot} s}
\label{beta}
\end{equation}

\noindent where $G$ is the gravitational constant, $R$ the heliocentric distance and $M_{\odot}$ the mass of the Sun, which gives $\beta$ = 2$\times$10$^{-4} ~u^2$. The characteristic travel time from the nucleus to the apex of motion is $\tau$ = $2s/u$. Assuming that the observed dust is on average 2 months old (as inferred from the tail position angles), we obtain ejection speeds of $u$ = 0.8\, m s$^{-1}$, and hence a typical $\beta$ = 1.3$\times$10$^{-4}$ (radius of 5\,mm).

\begin{figure}
\epsscale{0.85}
\begin{center}
\includegraphics[width=0.95\textwidth, angle =0 ]{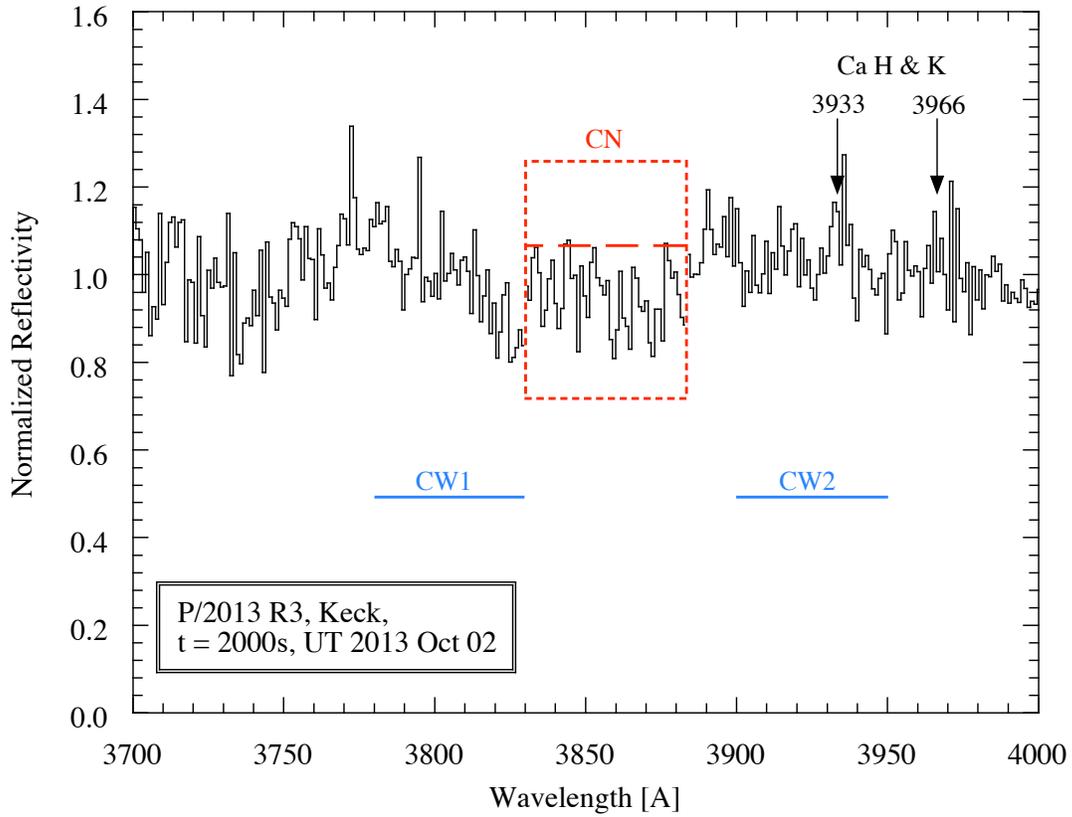}

\caption{R-band photometry from KAIT (black circles) and from Keck and HST (orange diamonds) within a 6\arcsec~radius circular aperture as a function of Day of Year in 2013.   Lines show the brightness variation expected of an asteroid following the inverse square law of distance and with phase functions appropriate for C-type (blue) and S-type (red) surfaces.   \label{KAIT}
} 
\end{center} 
\end{figure}

Integrated light photometry was extracted from KAIT data using a 6\arcsec~radius circular aperture centered on the optocenter,
with calibration from field stars (Figure \ref{KAIT}). Curves in the Figure show the  brightness variation expected from the changing observing geometry (Equation \ref{inversesq}).  We plot two estimates of $\Phi(\alpha)$, applicable to C-type (blue line; low albedo, primitive composition) and S-type (red line; high albedo, thermally metamorphosed composition) asteroids.  Figure \ref{KAIT} shows that the apparent fading of R3 by $\sim$2 mag~is largely a result of the changing observational geometry and that the dust cross-section remains nearly constant from October to December.  Near-constancy of the cross-section over 2.5 months  is consistent with a large mean particle size in the coma.

The KAIT photometry corresponds to an effective radius  $r_e \sim$ 2.6 to 3.0 km (Table \ref{photometry}), and to cross-section $C_e = \pi r_e^2$ = 21 to 28 km$^2$.  As may be seen by comparing the 6\arcsec~vs.~the 0.2\arcsec~photometry in Table \ref{photometry}, almost all of this cross-section lies in coma dust structures in R3, not in the embedded nuclei.  The mass and cross-section of an optically thin dust cloud are related by $M_d \sim 4/3 \rho \overline{a} C_e$, where $\rho$ is the dust mass density and $\overline{a}$ is the weighted mean dust grain radius.  For $\overline{a} \gtrsim$ 5 mm and $\rho$ = 10$^3$ kg m$^{-3}$ we estimate peak dust masses (on Oct 29) $M_d \gtrsim$ 2$\times$10$^8$ kg, equivalent to a $\sim$35 m radius  sphere having the same density.

\section{DISCUSSION}
Break-up of cometary nuclei has been frequently observed (Boehnhardt 2004) and variously interpreted as due to tidal stresses (Asphaug and Benz 1996), the build up of internal pressure forces from gases generated by sublimation (Samarasinha 2001), impact (Toth 2001) and rotational bursting (Jewitt 1992).    

The orbit of R3 (perihelion 2.20 AU, aphelion 3.8 AU) prevents close approaches to the sun or planets, so that tidal forces can be ignored.  To estimate the highest possible gas pressure on R3 we solved the energy balance equation for  black ice sublimating at the subsolar point.  The resulting equilibrium temperature, $T_{SS}$ = 197 K at 2.25 AU, corresponds to gas pressure  $P \sim$ 0.04 N m$^{-2}$, which is far smaller than both the central hydrostatic pressure and the $\sim$10$^3$ N m$^{-2}$ tensile strengths of even highly porous dust aggregates (Blum and Schr{\"a}pler~2004, Meisner et al.~2012, Seizinger et al.~2013). A more volatile ice (e.g.~CO), if present, could generate higher pressures but the long term stability of such material in the asteroid belt seems highly improbable. We conclude that sublimation gas pressure cracking is not a viable mechanism, although, if ice does exist in R3, its exposure after break-up could contribute to the continued dust production.

Several observations argue against an impact origin.  The separation times of the components are staggered over several months, whereas impact should give a single time.   Ejecta from an impact should be consistent with a single synchrone date whereas in R3 the fitted dates differ.  The scattering cross-section increases between October 01 and 29 and decreases very slowly thereafter (Table \ref{photometry}),  inconsistent with an impulsive origin. 
and unlike the best-established asteroid impact event (on (596) Scheila, c.f.~Bodewits et al.~2011, Jewitt et al.~2011,  Ishiguro et al.~2011).  
Furthermore,  impacts produce ejecta with a broad spectrum of velocities, from sub-escape to the impact speed (Housen and Holsapple 2011) whereas our data provide no evidence for fast ejecta, even in the earliest observations.   
For these reasons, we suspect that impact does not provide a natural explanation of the properties of R3, although we cannot rule it out.

Rotational breakup of a strengthless body should occur when the centripetal acceleration on the surface exceeds the gravitational acceleration towards the center.  For a sphere of density $\rho$ = 10$^3$ kg m$^{-3}$ the critical period for breakup is $\sim$3.3 hr, while for elongated bodies, the instability  occurs at longer periods.   Solar radiation provides a torque  (the ``YORP'' torque) capable of driving the spin of a sub-kilometer asteroid to the critical value in less than a million years, making rotational breakup a plausible mechanism for R3 and other small asteroids (Marzari et al.~2011).  (A tangential jet from sublimating ice carrying 1 kg s$^{-1}$ (i.e.~satisfying our spectral upper limit) could spin-up a 200 m radius body on a timescale of months).  Aspects of R3 consistent with rotational breakup include the absence of fast ejecta, the low velocity dispersion of the major fragments (comparable to the gravitational escape speeds)  and their peculiar alignment  (along the ABC axis in Figure \ref{images}), which we interpret as the rotational equator of the disrupted parent body.   Rotational instability is a potential source of bound (e.g.~Walsh et al.~2012) and unbound asteroid pairs (Jacobson et al.~2014, Polishook et al.~2014) and of chaotic systems in which mass is both re-accreted and shed from interacting ejecta (Jacobson and Scheeres 2011).  Six-tailed object P/2013 P5 has been interpreted as the product of rotational instability, although its morphology is quite different from that of R3 (Jewitt et al.~2013).  Depending on the body shape and material properties, the criteria for shedding instability and structural failure can be quite different (Hirabayashi and Scheeres 2014).  We suggest that P/2013 P5 is episodically shedding only its regolith while the multiple components of R3 indicate that a more profound structural failure has occurred.

Fresh observational effort is warranted to secure additional high-resolution measurements of the motions of the fragments in order to better constrain the dynamics of R3. Continued physical observations are also needed to isolate the embedded nuclei, and so to determine their sizes, shapes and rotational states.

\clearpage

\section{SUMMARY}

The main properties of active asteroid P/2013 R3, deduced from data taken  between UT 2013 October 01 and December 13,  are:

\begin{enumerate}

\item Asteroid P/2013 R3 is   split into at least 10 fragments, the largest of which have effective radii $\lesssim$ 200 m (geometric albedo 0.05 assumed).
The  fragments exhibit a velocity dispersion $\sim$ 0.2 to 0.5 m s$^{-1}$  and their motions indicate break-up dates in the range 2013 February to September.

\item The enveloping debris  cloud has integrated cross-section 21 to 29 km$^2$, an effective particle radius $\sim$5 mm, a total dust mass $M_d \sim$ 2$\times$10$^{8}$ kg and was ejected over a period of months.  The characteristic dust speed is $\sim$ 1 m s$^{-1}$. While the integrated cross-section deduced from photometry  is nearly constant, individual fragments fade at up to 1 mag.~month$^{-1}$.

\item  The spectrum  consists of sunlight reflected from dust, with no evidence for comet-like outgassing and a limit to water production $<$1 kg s$^{-1}$.  The optical colors are consistent with classification as a primitive C-type body.

\item The small velocity dispersion, staggered separation dates and initially linear arrangement of similarly-sized fragments lead us to suspect that P/2013 R3 is undergoing a rotationally triggered disruption.

\end{enumerate}

\acknowledgments
We thank Chris Snead for assistance, Masateru Ishiguro for discussions, Scott Sheppard for Magellan data and the anonymous referee for comments.  Based in part on observations made with the NASA/ESA \emph{Hubble Space Telescope,} with data obtained at the Space Telescope Science Institute (STSCI).  Support for program 13612 was provided by NASA through a grant from STSCI, operated by AURA, Inc., under contract NAS 5-26555.  We thank Linda Dressel, Alison Vick and other members of the STScI ground system team for their expert help.   Some of the data presented herein were obtained at the W.M. Keck Observatory, operated as a scientific partnership among Caltech, the University of California and NASA. The Observatory was made possible by the generous financial support of the W. M. Keck Foundation.

\clearpage

\begin{deluxetable}{lcccccccc}
\tablecaption{Observing Geometry 
\label{geometry}}
\tablewidth{0pt}
\tablehead{ \colhead{UT Date and Time} & DOY &
\colhead{Tel\tablenotemark{a}}  & \colhead{$R$\tablenotemark{b}}  & \colhead{$\Delta$\tablenotemark{c}} & \colhead{$\alpha$\tablenotemark{d}}   & \colhead{$\theta_{\odot}$\tablenotemark{e}} &   \colhead{$\theta_{-v}$\tablenotemark{f}}  & \colhead{$\delta_{\oplus}$\tablenotemark{g}}   }
\startdata
2013 Oct 01  07:45 - 08:20 &   274 & Keck  &  2.231 & 1.231  & 1.65 & 235.1 & 246.2  & -0.3\\
2013 Oct 29  06:36 - 08:17 &   302 &   HST  &  2.262 & 1.340  & 12.1 & 68.4 & 245.8  & -0.5\\
2013 Nov 15   06:39 - 07:20  &   319 &   HST   & 2.287 & 1.489  & 18.2 & 67.6 & 245.7  & -0.6\\
2013 Dec 13   07:25 - 08:05   & 347   &   HST  & 2.336 & 1.827 & 23.5 & 67.2 & 245.9 & -0.5 \\

\enddata


\tablenotetext{a}{Telescope used}
\tablenotetext{b}{Heliocentric distance, in AU}
\tablenotetext{c}{Geocentric distance, in AU}
\tablenotetext{d}{Phase angle, in degrees}
\tablenotetext{e}{Position angle of the projected anti-Solar direction, in degrees}
\tablenotetext{f}{Position angle of the projected negative heliocentric velocity vector, in degrees}
\tablenotetext{g}{Angle of Earth above the orbital plane, in degrees}

\end{deluxetable}

\clearpage

\begin{deluxetable}{lccccc}
\tablecaption{Nucleus Photometry\tablenotemark{a} 
\label{photometry}}
\tablewidth{0pt}
\tablehead{
\colhead{Name}         & \colhead{Oct 01} & \colhead{Oct 29} & \colhead{Nov 15} & \colhead{Dec 13}\\
\colhead{}         & \colhead{$V$~~~~$H_V$~~~~$r_e$ } & \colhead{$V$~~~~$H_V$~~~~$r_e$} & \colhead{$V$~~~~$H_V$~~~~$r_e$}  & \colhead{$V$~~~~$H_V$~~~~$r_e$}}

\startdata
A1 &  --  & 23.05 (19.91) 0.32  & 24.12 (20.63) 0.23 & 24.98 (20.85) 0.21 \\
A2  & --  & 23.72 (20.58) 0.24  & 24.16 (20.67) 0.23  & 25.09 (20.96) 0.20  \\
B\tablenotemark{b}     & -- & 22.04 (18.90) 0.51  & 23.47 (19.98) 0.31   & -- \\
B1 & -- & --  & --   & 24.97 (20.84) 0.21\\
B2 & -- & --  & --   & 25.18 (21.05) 0.19\\
C1   &  -- & 26.45 (23.31) 0.07  & $\ge$26.8 ($\ge$23.6) $\le$0.06   & $\ge$26.8 ($\ge$22.7) $\le$0.09\\
C2  &  -- & 24.86 (21.72) 0.14  & 25.35 (21.86) 0.13  & 26.37 (22.24) 0.11\\
Total &  17.81 (15.41) 2.57 & 18.19 (15.05) 3.02 & 18.72 (15.23) 2.78 & 19.46 (15.34) 2.64 \\

\enddata


\tablenotetext{a}{Three quantities are listed for each feature; $V$, the apparent magnitude,  $H_V$, the absolute magnitude computed from Equation (\ref{inversesq}) and $r_e$ [km], the effective radius  (Equation \ref{rn}).  For each component, we measured $V$ within a 0.2\arcsec~radius aperture with background subtraction from a contiguous annulus extending to 0.4\arcsec.  For the ``Total'' light measurement, we used a 6.0\arcsec~radius aperture with background annulus outer radius 12.0\arcsec.}
\tablenotetext{b}{The components of ``B'' were not separately measurable in data taken before December 13.}

\end{deluxetable}

\clearpage

\clearpage

\clearpage

\clearpage

\clearpage

\begin{thebibliography}{}

\bibitem[A'Hearn et al.(1995)]{1995Icar..118..223A} A'Hearn, M.~F., Millis, R.~C., Schleicher, D.~O., Osip, D.~J., \& Birch, P.~V.\ 1995, \icarus, 118, 223 

\bibitem[Asphaug \& Benz(1996)]{1996Icar..121..225A} Asphaug, E., \& Benz, W.\ 1996, \icarus, 121, 225 

\bibitem[Blum \& Schr{\"a}pler(2004)]{2004PhRvL..93k5503B} Blum, J., \& Schr{\"a}pler, R.\ 2004, Physical Review Letters, 93, 115503 

\bibitem[Bodewits et al.(2011)]{2011ApJ...733L...3B} Bodewits, D., Kelley, M.~S., Li, J.-Y., et al.\ 2011, \apjl, 733, L3 

\bibitem[Boehnhardt(2004)]{2004come.book..301B} Boehnhardt, H.\ 2004, Comets II, Univ. Arizona Press, Tucson, p.301 

\bibitem[Bowell et al.(1989)]{1989aste.conf..524B} Bowell, E., Hapke, B., Domingue, D., et al.\ 1989, in Asteroids II, Tucson, AZ, University of Arizona Press, p. 524-556. 

\bibitem[DeMeo \& Carry(2013)]{2013Icar..226..723D} DeMeo, F.~E., \& Carry, B.\ 2013, \icarus, 226, 723 

\bibitem[Dressel(2012)]{2012wfci.book.....D} Dressel, L.\ 2012, Wide Field Camera 3, HST Instrument Handbook,  

\bibitem[Finson \& Probstein(1968)]{1968ApJ...154..327F} Finson, M.~J., \& Probstein, R.~F.\ 1968, \apj, 154, 327

\bibitem[Hill et al.(2013)]{2013CBET.3658....1H} Hill, R.~E., Armstrong, J.~D., Molina, M.,  \& Sato, H.\ 2013, Central Bureau Electronic Telegrams, 3658, 1 

\bibitem[Hirabayashi \& Scheeres(2014)]{2014ApJ...780..160H} Hirabayashi, M., \& Scheeres, D.~J.\ 2014, \apj, 780, 160 

\bibitem[Holmberg et al.(2006)]{2006MNRAS.367..449H} Holmberg, J., Flynn, C., \& Portinari, L.\ 2006, \mnras, 367, 449 

\bibitem[Housen \& Holsapple(2011)]{2011Icar..211..856H} Housen, K.~R., \& Holsapple, K.~A.\ 2011, \icarus, 211, 856 

\bibitem[Hsieh \& Jewitt(2006)]{2006Sci...312..561H} Hsieh, H.~H., \& Jewitt, D.\ 2006, Science, 312, 561 

\bibitem[Ishiguro et al.(2009)]{2009Icar..203..560I} Ishiguro, M., Usui, F., Sarugaku, Y., \& Ueno, M.\ 2009, \icarus, 203, 560 

\bibitem[Ishiguro et al.(2011)]{2011ApJ...741L..24I} Ishiguro, M., Hanayama, H., Hasegawa, S., et al.\ 2011, \apjl, 741, L24 

\bibitem[Jacobson \& Scheeres(2011)]{2011Icar..214..161J} Jacobson, S.~A., \& Scheeres, D.~J.\ 2011, \icarus, 214, 161 

\bibitem[Jacobson et al.(2014)]{2014arXiv1401.1813J} Jacobson, S.~A., Marzari, F., Rossi, A., Scheeres, D.~J., \& Davis, D.~R.\ 2014, arXiv:1401.1813 

\bibitem[Jewitt(1992)]{1992LIACo..30...85J} Jewitt, D.\ 1992, Liege International Astrophysical Colloquia, 30, 85 

\bibitem[Jewitt(2012)]{2012AJ....143...66J} Jewitt, D.\ 2012, \aj, 143, 66 

\bibitem[Jewitt et al.(2011)]{2011ApJ...733L...4J} Jewitt, D., Weaver, H., Mutchler, M., Larson, S., \& Agarwal, J.\ 2011, \apjl, 733, L4 

\bibitem[Jewitt et al.(2013)]{2013ApJ...778L..21J} Jewitt, D., Agarwal, J., Weaver, H., Mutchler, M., \& Larson, S.\ 2013, \apjl, 778, L21 

\bibitem[Kresak(1980)]{1980M&P....22...83K} Kresak, L.\ 1980, Moon and Planets, 22, 83 

\bibitem[Landolt(1992)]{1992AJ....104..340L} Landolt, A.~U.\ 1992, \aj, 104, 340 

\bibitem[Marzari et al.(2011)]{2011Icar..214..622M} Marzari, F., Rossi, A., \& Scheeres, D.~J.\ 2011, \icarus, 214, 622 

\bibitem[Meisner et al.(2012)]{2012A&A...544A.138M} Meisner, T., Wurm, G., \& Teiser, J.\ 2012, \aap, 544, A138 

\bibitem[Oke et al.(1995)]{1995PASP..107..375O} Oke, J.~B., Cohen, J.~G., Carr, M., et al.\ 1995, \pasp, 107, 375 

\bibitem[Polishook et al.(2014)]{2014arXiv1401.4465P} Polishook, D., Moskovitz, N., Binzel, R.~P., et al.\ 2014, arXiv:1401.4465 

\bibitem[Reach et al.(2009)]{2009Icar..203..571R} Reach, W.~T., Vaubaillon, J., Kelley, M.~S., Lisse, C.~M., \& Sykes, M.~V.\ 2009, \icarus, 203, 571 

\bibitem[Richmond et al.(1993)]{1993PASP..105.1164R} Richmond, M., Treffers, R.~R., \& Filippenko, A.~V.\ 1993, \pasp, 105, 1164 

\bibitem[Samarasinha(2001)]{2001Icar..154..540S} Samarasinha, N.~H.\ 2001, \icarus, 154, 540 

\bibitem[Schleicher(2010)]{2010AJ....140..973S} Schleicher, D.~G.\ 2010, \aj, 140, 973 

\bibitem[Seizinger et al.(2013)]{2013A&A...559A..19S} Seizinger, A., Speith, R., \& Kley, W.\ 2013, \aap, 559, A19 

\bibitem[Toth(2001)]{2001A&A...368L..25T} Toth, I.\ 2001, \aap, 368, L25 


\bibitem[Walsh et al.(2012)]{2012Icar..220..514W} Walsh, K.~J., Richardson, D.~C., \& Michel, P.\ 2012, \icarus, 220, 514 


\bibitem[Weaver et al.(2001)]{2001Sci...292.1329W} Weaver, H.~A., Sekanina, Z., Toth, I., et al.\ 2001, Science, 292, 1329 

\bibitem[Weaver et al.(2008)]{2008LPICo1405.8248W} Weaver, H.~A., Lisse, C.~M., Mutchler, M., et al.\ 2008, LPI Contributions, 1405, 8248 

\end{thebibliography}
\end{document}